\documentclass[journal,twoside]{IEEEtran}

\usepackage{amsmath}
\usepackage{amssymb}
\usepackage{bm}
\usepackage{graphicx}
\usepackage{cite}
\usepackage{algorithmic}
\usepackage{algorithm}

\newcommand{\bH}{\ensuremath{\bm{H}}}

\newcommand{\bW}{\ensuremath{\bm{W}}}

\newcommand{\br}{\ensuremath{\bm{r}}}
\newcommand{\bv}{\ensuremath{\bm{v}}}
\newcommand{\bn}{\ensuremath{\bm{n}}}
\newcommand{\varn}{\ensuremath{\sigma_{n}^{2}}}
\newcommand{\bI}{\ensuremath{\bm{I}}}
\newcommand{\sV}{\ensuremath{\mathcal{V}}}
\newcommand{\B}{\ensuremath{\bar{B}}}
\newcommand{\Nr}{\ensuremath{\bar{N}_r}}
\newcommand{\Cbt}{\ensuremath{\bar{R}_{\triangle}}}

\newcommand{\srvq}{\ensuremath{\gamma_{\infty}}}
\newcommand{\me}{\ensuremath{\mathrm{e}}}
\newcommand{\bvh}{\ensuremath{\hat{\bm{v}}}}
\newcommand{\diff}{\ensuremath{\mathrm{d}}}
\newcommand{\bL}{\ensuremath{\bm{\Lambda}}}

\newtheorem{lemma}{Lemma}

\DeclareMathOperator{\Tr}{tr}

\begin{document}

\title{On Optimizing Feedback Interval for Temporally Correlated MIMO Channels
  With Transmit Beamforming And Finite-Rate Feedback}

\author{Kritsada Mamat and Wiroonsak Santipach,~\IEEEmembership{Senior~Member,~IEEE}%
\thanks{This work was supported by postdoctoral funding from the Faculty of Engineering, Kasetsart University, Bangkok, Thailand under grant number 59/02/EE and by Kasetsart University Research and Development Institute (KURDI) under the FY2018 Kasetsart University research grant.}%
\thanks{The material in this paper was presented in part at the IEEE Global Communications Conference (GLOBECOM), Houston, Texas, USA, Dec. 2011~\cite{gc2011}.}%
\thanks{K. Mamat was with the Department of Electrical Engineering;
  Faculty of Engineering; Kasetsart University, Bangkok, 10900,
  Thailand. He is currently with the Department of Electronic Engineering Technology; College of Industrial Technology; King Mongkut's University of Technology North Bangkok, Bangkok, 10800, Thailand (email: kritsada.m@cit.kmutnb.ac.th).}
\thanks{W. Santipach is with the Department of Electrical Engineering;
  Faculty of Engineering; Kasetsart University, Bangkok, 10900,
  Thailand (email: wiroonsak.s@ku.ac.th).}}

\markboth{IEEE Transactions on Communications, 2018}{Mamat and Santipach: On
  Optimizing Feedback Interval for Temporally Correlated MIMO Channels}

\maketitle

\begin{abstract}
A receiver with perfect channel state information (CSI) in a
point-to-point multiple-input multiple-output (MIMO) channel can
compute the transmit beamforming vector that maximizes the
transmission rate.  For frequency-division duplex, a transmitter is
not able to estimate CSI directly and has to obtain a quantized
transmit beamforming vector from the receiver via a rate-limited
feedback channel.  We assume that time evolution of MIMO channels is
modeled as a Gauss-Markov process parameterized by a
temporal-correlation coefficient.  Since feedback rate is usually low,
we assume rank-one transmit beamforming or transmission with single
data stream.  For given feedback rate, we analyze the optimal
feedback interval that maximizes the average received power of the
systems with two transmit or two receive antennas.  For other system
sizes, the optimal feedback interval is approximated by maximizing the
rate difference in a large system limit.  Numerical results show that
the large system approximation can predict the optimal interval for
finite-size system quite accurately.  Numerical results also show that
quantizing transmit beamforming with the optimal feedback interval
gives larger rate than the existing Kalman-filter scheme does by as
much as 10\% and than feeding back for every block does by 44\% when
the number of feedback bits is small.
\end{abstract}

\begin{IEEEkeywords}
MIMO, transmit beamforming, temporally correlated channels,
Gauss-Markov process, finite-rate feedback, random vector quantization
(RVQ), feedback interval.
\end{IEEEkeywords}

\section{Introduction}

Employing multiple antennas at transmitters and/or receivers has been
shown to increase spatial diversity and spectral
efficiency~\cite{telatar,foschini98}.  To achieve higher potential of
multiple antennas, some channel state information (CSI) at both the
transmitter and receiver is required.  At a receiver, CSI can be
estimated from pilot signals.  However, estimating the channel at a
transmitter is not possible for frequency-division duplex (FDD) where
forward and backward channels are in different frequency bands.
Consequently, a transmitter in FDD must obtain CSI from a receiver via
a low-rate feedback channel.  Many researchers have proposed schemes
to quantize and feed back CSI and analyze the associated performance
(see \cite{love08} and references therein).  With finite feedback
rate, the beamforming vector is selected from a quantization set or a
codebook, which is known {\em a priori} at the transmitter and the
receiver.  The codebook index of the selected vector is then fed back
to the transmitter, which subsequently adjusts its beamforming
coefficients accordingly.  Different codebooks have been proposed and
analyzed in~\cite{love03,mimo,wcom11,ryan09}.  The optimal
Grassmannian codebook that maximizes the minimum chordal distance
between any two codebook entries was proposed
in~\cite{love03}. In~\cite{mimo}, a random vector quantization (RVQ)
codebook whose entries are independent isotropically distributed, is
analyzed.  RVQ codebook is simpler to construct than Grassmannian
codebook and performs close to the optimum.  To reduce search
complexity of RVQ, the codebook entries are organized in a tree
structure in~\cite{wcom11}.  In~\cite{ryan09}, PSK and QAM codebooks
were proposed with low-complexity search based on noncoherent
detection algorithm.  If CSI at the receiver is also not perfect due
to limited channel training, the rate performance will degrade
further.  Imperfect CSI at the receiver in conjunction with limited
feedback has been considered in our previous work~\cite{train10}.

Feeding back quantized beamforming coefficients may not be useful in a
fast fading channel since they are quickly outdated~\cite{ma09}.  If
the channel fades slowly, the beamforming coefficients may not need to
be updated frequently.  Thus, the feedback scheme should be adapted to
temporal correlation of the channel~\cite{mondal06, huang09, kim11,
  osmane13, zhang12, kim11dif, medra15}.  Switched codebook
quantization was proposed in~\cite{mondal06} where the codebook
selection was based on channel spatial and temporal correlations.
In~\cite{huang09}, quantized CSI was modeled as a first-order
finite-state Markov chain and beamforming feedback is based on the
channel dynamics.  An adaptive feedback period (AFP) scheme in which
the receiver feeds back to the transmitter periodically was considered
in~\cite{kim11}.  However, the authors were only concerned with MISO
channels in which the number of receive antennas is fixed to 1.  The
optimal feedback period for coordinated multi-point (COMP) systems was
considered in~\cite{osmane13} where channels are also modeled as a
first-order Gauss-Markov process.  In~\cite{zhang12}, the minimum
feedback rate of a differential feedback scheme was analyzed.  The
authors in~\cite{kim11dif} have proposed a differential codebook,
which is rotated according to channel correlation, feedback rate, and
the previous transmit beamforming. In~\cite{chen17,medra15}, a
differential precoder, which depends on temporal correlation of the
channel, adjusts the quantized transmit precoder to be closer to the
optimal precoder.

Another line of work~\cite{xu14, mehanna14, noh16} applied Kalman
filter (KF) to predict the current transmission channel based on
previous estimates and channel correlation.
References~\cite{xu14,mehanna14} proposed quantizing and feeding back
an innovation term, which is the difference between the received
signal and its estimate, to the transmitter.  The current channel
estimate then can be computed by the transmitter using KF with a
sequence of the previous quantized innovations.  In~\cite{mehanna14},
only 2 bits per update were required to send back innovations and were
used to compute the beamforming vector by the transmitter.  CSI at the
receiver was obtained via a pilot signal and was not perfect.
Reference~\cite{noh16} improved the training phase of KF beamforming
in massive MIMO systems by reducing the amount of pilot.

For this work, we consider block Rayleigh-fading MIMO channels with
time evolution modeled by a first-order Gauss-Markov process.  (An
uncorrelated block-fading model was considered in our previous
work~\cite{mimo, train10}).  Antennas are assumed to be sufficiently
far apart that they are independent.  We analyze the performance of
quantized beamforming (rank-one precoding) in the AFP scheme first
proposed by~\cite{kim11}, which considered only MISO channels.  In our
previous work~\cite{tc15}, we also considered quantizing transmit
beamforming in MISO channels, but in conjunction with orthogonal
frequency-division multiplexing (OFDM), and optimize the size of
subcarrier cluster.  To quantize transmit beamforming, we apply random
vector quantization (RVQ) codebook, which has been shown to perform
close to the optimum codebook~\cite{mimo,commag04}.  Furthermore, RVQ
can be analyzed to obtain some insights into the limited feedback
performance.  Although transmission with beamforming or rank-one
precoding does not achieve full spatial multiplexing gain in MIMO
channels, the amount of CSI feedback required for beamforming is
substantially smaller than that with full-rank precoding~\cite{mimo}.
As subsequent results will show, the AFP scheme with our proposed
feedback interval outperforms other schemes in low-feedback regimes.
Also, when feedback rate is low, the optimal rank of the precoding
matrix that maximizes achievable rate is also low and thus, transmit
beamforming can be optimal or close to optimal~\cite{mimo}.  Hence, our
contribution, which is stemmed from quantizing transmit beamforming,
will be most beneficial for systems with very limited feedback.

In this study, we can summarize our contribution as follows
\begin{itemize}

\item We derive a closed-form expression of the averaged received
  power for channels with two transmit antennas and arbitrary number
  of receive antennas, which is based on the eigenvalue distribution
  of the channel matrix~\cite{veeravalli13}.  For channels with
  arbitrary number of transmit and two receive antennas, the
  expression for the averaged received power is also derived, but
  needs to be evaluated numerically.  We formulate the problems that
  find the optimal feedback interval and compare the rate performance
  of AFP scheme and the minimum feedback-period (MFP) scheme, which
  updates feedback for every fading block.  Similar study has been
  performed in~\cite{kim11} for MISO channels and in~\cite{osmane13}
  for COMP system with a single-antenna receiver.  However, our
  results, which apply to MIMO models as well, are different and not
  simple extension of~\cite{kim11} or~\cite{osmane13}.  We find that
  the maximum feedback interval where the AFP scheme outperforms the
  MFP one, depends more on the number of receive antennas especially
  when feedback rate is low.

\item For channels with an arbitrary number of transmit and receive
  antennas, we derive the averaged rate difference in a large system limit
  in which the numbers of transmit and receive antennas and the number
  of feedback bits tends to infinity with fixed ratios.  Numerical
  examples show that the large system results can be used to
  approximate the optimal feedback interval of finite-size systems.
  Some of the large system results were presented in part
  in~\cite{gc2011}.

\item Our numerical results show that the AFP scheme with the optimal
  feedback interval outperforms KF beamforming with quantized
  innovation in all feedback-rate regimes and the performance gain can
  be significant in MIMO channels.  We also find that with very low
  feedback rate, the AFP scheme achieves larger averaged received
  power than the differential codebook proposed by~\cite{kim11dif},
  which is adapted with the channel.  Although the optimal feedback
  interval is analyzed for RVQ codebook, the numerical results show
  that the optimal feedback interval for RVQ is close to that for
  Grassmannian codebook, which achieves optimal rate for channels with
  finite number of antennas.
\end{itemize}

The paper is organized as follows.  Section~\ref{sys_mod} introduces
the channel model and feedback schemes.  In Section~\ref{afp}, we
analyze the optimal feedback interval for systems with two transmit
and/or two receive antennas.  Large system analysis is shown in
Section~\ref{large_sys}.  The numerical results and conclusions are in
Sections~\ref{num_re} and~\ref{conclude}, respectively.

\section{System Model}
\label{sys_mod}

We consider a point-to-point discrete-time multiple-antenna channel with
$N_t$ transmit and $N_r$ receive antennas.  We assume block fading in
which the channel gains remain static for $L$ symbols and change in the
next block of symbols.  To allow meaningful feedback of CSI from a
receiver, the block length $L$, which is also a coherence period, is
assumed to be sufficiently long.  During the $k$th fading block, an
$N_r \times 1$ receive vector during symbol index $kL + l$ is given by
\begin{equation}
  \br[kL + l] = \bH(k) \bv(k) x_s[kL + l] + \bn[kL + l], \quad 1 \le l
  \le L
\label{eq_kL}
\end{equation}
where we use square brackets and parentheses to indicate symbol index
and block index, respectively.  In~\eqref{eq_kL}, $x_s[i]$ is the
$i$th transmitted symbol with zero mean and unit variance, $\bn[i]$ is
an $N_r \times 1$ additive white Gaussian noise (AWGN) vector during
symbol index $i$ with zero mean and covariance $\varn \bI$ where $\bI$
is an identity matrix, $\bv(k)$ is an $N_t \times 1$ unit-norm
beamforming vector for the $k$th fading block, and $\bH(k) =
[h_{ij}(k)]$ is an $N_r \times N_t$ channel matrix whose element
$h_{ij}(k)$ is the channel gain between the $i$th receive and the
$j$th transmit antennas during the $k$th fading block.  Here, we
consider rank-one transmit precoding or beamforming.  Arbitrary-rank
transmit precoding with multiple independent data streams in
temporally uncorrelated MIMO channels was considered in~\cite{mimo}.
Assuming an ideal scattering environment, $h_{ij}(k)$ is modeled as a
complex Gaussian random variable with zero mean and unit variance.
Also, we assume that adjacent antennas in antenna arrays at both the
transmitter and receiver are placed sufficiently far apart that
elements of $\bH(k)$ are independent.

To model a time evolution of the channel considered, we adopt the
first-order Gauss-Markov process, which has been widely used for its
tractability~\cite{zhao_pimrc05,mondal06,peel07,kim11}.  Thus, the
channel matrix of the $k$th fading block relates to that of the
previous block as follows
\begin{equation}
    \bH(k) = \alpha \bH(k-1) + \sqrt{1 - \alpha^2} \bW(k)
\label{eq_1od}
\end{equation}
where $\bW(k)$ is an $N_r \times N_t$ innovation matrix with
independent zero-mean unit-variance complex Gaussian entries, and
$\alpha \in [0,1)$ denotes a temporal correlation coefficient between
  adjacent blocks.  Note that $\alpha \rightarrow 1$ produces a
  time-invariant channel.  On the other hand, $\alpha = 0$ indicates a
  channel with no temporal correlation and thus, the channel fades
  independently from one coherence block to the next.  For the
  Jakes/Clarke fading model~\cite{jake93}, $\alpha = \mathbb{J}_0(2
  \pi D_s T_s)$ where $\mathbb{J}_0(\cdot)$ is the zeroth-order Bessel
  function, $D_s$ is the Doppler spread, and $T_s$ is the time
  duration of a block.  For example, for a channel with 900-MHz
  carrier frequency and 5-ms average fading block, $\alpha$ ranges
  from 0.5 to 0.9999 as mobile's velocity varies from 60 km/h to 1
  km/h.

The associated ergodic achievable rate of this channel is given by
\begin{equation}
  R = E \left[ \log \left(1 + \rho \bv(k)^\dag \bH(k)^\dag \bH (k) \bv
    (k) \right) \right]
  \label{eq_Ell}
\end{equation}
where $\rho = E[|x_s|^2]/\varn = 1/\varn$ denotes the background
signal-to-noise ratio (SNR), $[\cdot]^{\dag}$ denotes the Hermitian
transpose, and $E[\cdot]$ denotes the expectation operator.  We note
that the expectation in~\eqref{eq_Ell} is over channel matrix.  To
achieve the desired rate, the transmitter encodes the transmitted
symbols across many different fading blocks with equal power per
symbol.  In addition to SNR, the achievable rate also depends on the
beamforming vector $\bv(k)$.  If the transmitter can track the channel
perfectly (perfect CSI), the optimal $\bv(k)$ is the eigenvector of
$\bH(k)^{\dag} \bH(k)$ corresponding to the maximum eigenvalue.  In
other words, the optimal beamforming vector is in the direction of the
strongest channel mode.

With FDD, the transmitter is not able to estimate the channel directly
and has to rely on CSI fed back from the receiver via a rate-limited
channel.  The receiver can estimate the channel from pilot signals,
which is known {\em a priori} at the transmitter and receiver.
Assuming perfect CSI, the receiver selects the optimal beamforming
vector and sends it back via a feedback channel to the transmitter.
Since the feedback channel is rate-limited, the selected beamforming
vector needs to be quantized.  Here, we quantize the transmit
beamforming vector with an RVQ codebook
\begin{equation}
  \sV = \{ \bv_1, \bv_2, \cdots, \bv_n\}
\end{equation}
where entries $\bv_j$ are independent isotropically distributed and
$n$ denotes the number of entries in the RVQ codebook.  For given
$\log_2 n$ quantization bits, RVQ performs close to the optimal
codebook~\cite{commag04,mimo} for channels with finite number of
transmit and receive antennas.  In a large system limit to be defined,
RVQ is optimal (i.e., maximizes achievable rate)~\cite{mimo,dai09}.

Given $\log_2 n$ bits and channel matrix $\bH (k)$, the receiver selects
from the RVQ codebook
\begin{align}
  \bvh(k) &= \arg \max_{\bv_j \in \sV} \ \log \left(1 + \rho
  \bv^{\dag}_j \bH(k)^{\dag} \bH (k) \bv_j \right)\\ &= \arg
  \max_{\bv_j \in \sV} \ \bv^{\dag}_j \bH(k)^{\dag} \bH (k) \bv_j .
\end{align}
The index of the selected beamforming vector is then fed back to the
transmitter, which adjusts its beamforming vector accordingly.  We
assume that the time duration to feed back the selected index is
negligible when compared to one fading block and that the
feedback channel is error-free.  The associated achievable rate with
a quantized transmit beamformer is given by
\begin{equation}
  R = E \left[ \log \left(1 + \rho \bvh(k)^{\dag} \bH(k)^{\dag} \bH
    (k) \bvh (k) \right) \right] .
  \label{c_rvq}
\end{equation}
Since the channel is time-varying, the transmit beamforming needs to
be quantized and fed back for every fading block.  This may not be
practical due to the limited feedback rate.  However, the system can
take advantage of temporal correlation of the channel in order to
reduce the number of bits needed.  In this paper, we consider feedback
schemes that reduce the number of feedback bits while maintaining
performance.

\section{On Optimizing Feedback Interval}
\label{afp}

Suppose that there are $B$ feedback bits available per fading block.
Since the overhead must be kept small, $B$ bits per fading block may
not be sufficient to meaningfully quantize a beamforming vector $\bv$.
In the AFP scheme proposed by~\cite{kim11}, $\bv$ is quantized and fed
back at the beginning of every interval of $K$ fading blocks with $BK$
bits instead of every block with $B$ bits.  However, the transmit
beamforming vector quantized to the first fading block with more
feedback bits will gradually be outdated as time passes.  Thus, the
feedback interval $K$ should be adjusted to the temporal correlation
of the channel.  In this section, we analyze the optimal feedback
interval for MIMO channels in the AFP scheme.  Note that the feedback
interval was analyzed for MISO channels by~\cite{kim11}.  Here we
analyze the achievable rate for MIMO channels with either two transmit
or two receive antennas.  The analysis involves the eigenvalue
distribution of the channel matrix and the distribution of the
received power with RVQ codebook conditioned on the
channel~\cite{veeravalli13}, which becomes more complex as the system
size increases.  Thus, our results are not simple extension of those
in~\cite{kim11}.

First, we determine an average achievable rate over $K$ fading blocks
given by
\begin{align}
   \bar{R} &= \frac{1}{K}\sum_{k=1}^{K}E\left[ \log \left(1 + \rho
     \bvh(1)^{\dag} \bH(k)^{\dag} \bH (k) \bvh (1)
     \right)\right] \label{c_interval}\\
   &\leq \frac{1}{K}\sum_{k=1}^{K} \log \left(1 + \rho
     E\left[\bvh(1)^{\dag} \bH(k)^{\dag} \bH (k) \bvh (1)\right]
     \right) \label{c_jensen}\\
   &\le \log \left(1 + \rho
   \frac{1}{K}\sum_{k=1}^{K} E \left[ \bvh(1)^{\dag} \bH(k)^{\dag} \bH
     (k) \bvh (1) \right] \right) \label{cbnd}
\end{align}
where $\bvh (1)$ is the quantized transmit beamformer for the channel
$\bH (1)$ in the first fading block and we apply Jensen's inequality
to obtain the upper bound~\eqref{cbnd}.  From~\eqref{c_interval}, we
see that for the AFP scheme, the quantized beamformer of the first
block is used for all $K$ consecutive blocks.  Since the expression of
the average rate in~\eqref{c_interval} is not tractable, we choose to
instead maximize the rate upper bound in~\eqref{cbnd} and obtain the
feedback interval as follows
\begin{equation}
  K^* = \arg \max_{K \in \mathbb{Z}^+} \frac{1}{K}\sum_{k=1}^{K}E
  \left[ \bvh(1)^{\dag} \bH(k)^{\dag} \bH (k) \bvh (1)\right],
\label{optK1}
\end{equation}
which is an integer optimization problem.  The problem
in~\eqref{optK1} is to maximize the average received power over $K$
blocks.  If $K$ is not too large, an exhaustive search can be
performed to find the optimal feedback interval $K^*$.  We expect
$K^*$ to be a good estimate of the feedback interval that maximizes
the average rate~\eqref{c_interval} in a low-SNR regime since in that
regime, logarithm increases approximately linearly with the received
power.

\subsection{$2 \times N_r$ Channels}
\label{sub:2r}

For a point-to-point channel with 2 transmit antennas and $N_r > 1$
receive antennas, the following lemma gives the expected received
power during the $k$th fading block when the quantized transmit
beamforming for the first block is used.

\begin{lemma}
\label{lemma_Nr}
The received power for the $k$th block of a $2 \times N_r$ channel
with $BK$ bits to quantize $\bv(1)$, is given by
\begin{multline}
  E\left[ \bvh(1)^{\dag} \bH(k)^{\dag} \bH(k) \bvh(1) \right] \\
  = \alpha^{2k-2} \left( \gamma_{2 \times N_r}(BK) - N_r \right) +
  N_r \label{GkNr}
\end{multline}
where
\begin{align}
   \gamma_{2 \times N_r}(BK) &\triangleq E\left[ \bvh(1)^{\dag}
     \bH(1)^{\dag} \bH(1) \bvh(1) \right]\\
  &= \frac{1}{(N_r-1)!(N_r-2)!} \Big[ \phi(N_r+2,N_r-1) \nonumber \\
    &\quad -2\phi(N_r+1,N_r) + \phi(N_r,N_r+1)\nonumber \\
    &\quad - \frac{1}{2^{BK}+1} \big(\phi(N_r+2,N_r-1) \nonumber \\
    &\quad - 3 \phi(N_r+1,N_r) + 3\phi(N_r,N_r+1) \nonumber \\
    &\quad  -\phi(N_r-1,N_r+2) \big) \Big] \label{GNr}
\end{align}
and $\phi(m,n)$ is a recursive function given by
\begin{multline}
  \phi(m,n) = mn\phi (m-1, n-1) - \frac{(m-n)(m+n-1)!}{2^{m+n+1}},\\
  \forall m, n \ge 1 \label{Rec1}
\end{multline}
with the following initial conditions: $\phi(4,1) = \frac{45}{8}$,
$\phi(3,2) = \frac{11}{8}$, $\phi(2,3) = \frac{5}{8}$, and $\phi(1,4)
= \frac{3}{8}$.
\end{lemma}
The proof is in Appendix~\ref{a1}.

From~\eqref{GkNr}, we see that as $k$ increases, the received power
decreases since the channel becomes less matched to the transmit
beamformer $\bvh(1)$.  However, if the channel is highly correlated
($\alpha$ close to 1), the received power will gradually decrease with
time.  Averaged over the whole feedback interval, the received power
for a $2 \times N_r$ channel is given by
\begin{multline}
   \frac{1}{K}\sum_{k=1}^{K} E\left[ \bvh^{\dag}(1)
     \bH^{\dag}(k)\bH(k)\bvh(1) \right]\\
   = N_r + \frac{1}{K} \left(
   \frac{1-\alpha^{2K}}{1-\alpha^2} \right)(\gamma_{2 \times N_r}(BK)
   - N_r) .\label{Ave2xNr}
\end{multline}
We note that the average received power increases with $B$.  To
determine $K^*$ that maximizes the average received power, we
substitute~\eqref{Ave2xNr} into~\eqref{optK1} and solve the problem.
To obtain some insight on $K^*$, we can consider the two extreme
regimes.  When channels are less correlated ($\alpha \to 0$) and $B$
is large, $K^*$ will be close to 1.  This is due to the diminishing
return of $\gamma_{2 \times N_r}(x)$. $K^* \approx 1$ implies that
feedback must occur as frequently as possible when the channel is fast
changing and feedback rate is high.  When channels are highly
correlated ($\alpha \to 1$), we can show with L'H\^{o}pital's rule
that
\begin{equation}
  \lim_{\alpha \to 1} \frac{1}{K}\sum_{k=1}^{K} E\left[ \bvh(1)^{\dag}
    \bH(k)^{\dag} \bH(k) \bvh(1) \right] = \gamma_{2 \times N_r}(BK) .
\label{AveNrx2lim}
\end{equation}
Thus, the optimal interval $K^* \to \infty$ since $\gamma_{2 \times
  N_r}(x)$ is increasing with $K$.  In other words, if the channel is
relatively static, the feedback interval should be large.  For other
values of $\alpha$ (e.g., $\alpha = 0.8$), our numerical results in
Fig.~\ref{OptNtNr} show that $K^*$ does not depend much on $N_r$ since
increasing the number of receive antennas seems to increase the
received signal power uniformly for all $K$.

In~\cite{kim11}, the performance of the AFP scheme is compared with
that of the minimum feedback period (MFP) scheme in which transmit
beamforming is quantized and fed back to the transmitter for every
fading block ($K = 1$).  However, \cite{kim11} only considers MISO
channels. In MIMO channels with a given feedback rate of $B$ bits per
fading block, we find that the AFP scheme (with $K > 1$) outperforms
the MFP scheme (with $K = 1$) if
\begin{equation}
  N_r + \frac{1}{K} \left( \frac{1-\alpha^{2K}}{1-\alpha^2} \right)
  (\gamma_{2 \times N_r}(BK) - N_r) > \gamma_{2 \times N_r}(B)
  \label{33b}
\end{equation}
where the right-hand side of~\eqref{33b} is the average received power
in \eqref{Ave2xNr} with $K = 1$. With some algebraic manipulation, we obtain
\begin{align}
  K &< \left(\frac{1 - \alpha^{2K}}{1 - \alpha^2}\right) \left(
  \frac{\gamma_{2 \times N_r}(BK) - N_r}{\gamma_{2 \times N_r}(B) -
    N_r} \right) \\
  &< \frac{1}{1 - \alpha^2} \left( \frac{\gamma_{2
      \times N_r}(BK) - N_r}{\gamma_{2 \times N_r}(B) - N_r} \right) .
\label{Klf}
\end{align}
Thus, \eqref{Klf} gives the range of $K$ in which the performance of
AFP exceeds that of MFP and the maximum $K$ with that property.  If we
consider a large $B$ regime or $B \to \infty$, the
inequality~\eqref{Klf} becomes
\begin{equation}
  K < \frac{1}{1 - \alpha^2} .
\end{equation}
Thus, we can conclude that when the feedback rate is large, the
maximum feedback interval of the AFP scheme that outperforms the MFP
scheme depends largely on the temporal correlation $\alpha$.  Thus,
the feedback interval for the AFP scheme can be set larger when
channels are highly correlated and should be shortened when channels
are less correlated.

\subsection{$N_t \times 2$ Channels}

Next, we consider channels with $N_t > 2$ transmit antennas and two
receive antennas.  We can follow the derivation of the averaged
received power for $2\times N_r$ channels in Section~\ref{sub:2r} to
obtain the averaged received power for $N_t \times 2$ channels,
\begin{multline}
   \frac{1}{K}\sum_{k=1}^{K} E\left[ \bvh^{\dag}(1)
     \bH^{\dag}(k)\bH(k)\bvh(1) \right]\\ = 2 + \frac{1}{K} \left(
   \frac{1-\alpha^{2K}}{1-\alpha^2} \right)(\gamma_{N_t \times 2}(BK)
   - 2)
\label{1a2}
\end{multline}
where the above expression follows~\eqref{Ave2xNr} with $N_r = 2$, and
\begin{equation}
  \gamma_{N_t \times 2}(BK) = E \left[ \bvh(1)^{\dag} \bH (1)^{\dag}
    \bH (1) \bvh(1)\right]
\end{equation}
is the received power of the first block.  Recall that
\begin{equation}
  \bvh(1)^{\dag} \bH (1)^{\dag} \bH (1) \bvh(1) = \max_{1 \le j \le
    2^{BK}} \bv_j^{\dag} \bH (1)^{\dag} \bH (1) \bv_j .
\label{maxv}
\end{equation}

Since the RVQ codebook is employed, the probability density function
(pdf) of $\bv_j^{\dag} \bH (1)^{\dag} \bH (1) \bv_j$ is identical for
all $j$ and is equal to $\bv_j^{\dag} \bL \bv_j$~\cite{veeravalli13}
where $\bL$ is an $N_t \times N_t$ diagonal matrix whose main diagonal
entries are the ordered eigenvalues of $\bH (1)^{\dag} \bH (1)$.  For
this channel, there are only two nonzero eigenvalues, which are
denoted by $\lambda_1$ and $\lambda_2$ and $\lambda_1 \ge \lambda_2 >
0$.  We derive the distribution of $\bv_j^{\dag} \bL \bv_j$ and obtain
the following lemma.

\begin{lemma}
\label{lemma_Nt}
Let $\bv$ be an $N_t \times 1$ isotropically distributed vector with
$N_t > 2$ and $\bL= \text{diag} ([\lambda_1, \lambda_2, \underbrace{0,
    0,..., 0}_{N_t-2}])$ with $\lambda_1 \geq \lambda_2 > 0$. The
cumulative distribution function (cdf) of $\bv^{\dag} \bL \bv$
conditioned on $\lambda_1$ and $\lambda_2$ is given by
\begin{multline}
F_{\bv^{\dag} \bL \bv |\lambda_1, \lambda_2}(x)\\ = \left\{%
\begin{array}{ll}
1-\frac{\lambda_1}{\lambda_1 - \lambda_2} \left( 1-\frac{x}{\lambda_1}
\right)^{N_t-1} & \\
\quad + \frac{\lambda_2}{\lambda_1 - \lambda_2} \left(
1-\frac{x}{\lambda_2} \right)^{N_t-1} & :\, 0\leq x \leq \lambda_2 \\
 1 - \frac{(\lambda_1-x)^{N_t-1}}{(\lambda_1-\lambda_2)\lambda_1^{N_t-2}} & :\, 
\lambda_2 \leq x \leq \lambda_1.
\end{array}%
\right.
\label{cdfNt}
\end{multline}
\end{lemma}
We remark that the expression of the cdf for $\lambda_2 \leq x \leq
\lambda_1$ is obtained from~\cite{veeravalli13} and is shown in
Lemma~\ref{lemma_Nt} for completeness. However, the expression of the
cdf for $0\leq x \leq \lambda_2$ is not derived in~\cite{veeravalli13}
and is not a simple extension of the earlier case.  The proof of
Lemma~\ref{lemma_Nt} is shown in Appendix~\ref{a2}.

With~\eqref{maxv} and \eqref{cdfNt}, it is straightforward to show
that
\begin{multline}
   E\left[ \bvh(1)^{\dag} \bH(1)^{\dag}\bH(1) \bvh(1) |\lambda_1,
     \lambda_2 \right]\\
   = \lambda_1 - \int_0^{\lambda_1} \left(
   F_{\bv^{\dag} \bL \bv}(x) \right)^{2^{BK}} \, \diff x .
\label{bvh}
\end{multline}

Thus,
\begin{multline}
E \left[ \bvh(1)^{\dag} \bH(1)^{\dag} \bH(1) \bvh(1) \right]\\ =
\int_{0}^{\infty} \int_{0}^{\lambda_1} E \left[ \bvh(1)^{\dag}
  \bH(1)^{\dag} \bH(1) \bvh(1) |\lambda_1,\lambda_2 \right]\\
\times f_{\bL}(\lambda_1,\lambda_2) \, \diff \lambda_1 \diff \lambda_2
\label{ld2}
\end{multline}
where $f_{\bL}(\lambda_1,\lambda_2)$ is the joint pdf of the two
ordered eigenvalues of $\bH(1)^{\dag} \bH(1)$ and is stated
in~\eqref{pdln} where $N_t$ replaces $N_r$.  Substitute~\eqref{bvh}
into~\eqref{ld2} and evaluate the first integral to obtain
\begin{multline}
  \gamma_{N_t \times 2}(BK) = \frac{\phi(N_t+2,
    N_t-1)}{(N_t-1)!(N_t-2)!}\\ - \int_{0}^{\infty}
  \int_{0}^{\lambda_1} \int_0^{\lambda_1} \left( F_{\bv^{\dag} \bL
    \bv}(x) \right)^{2^{BK}} f_{\bL}(\lambda_1,\lambda_2) \, \diff x
  \, \diff \lambda_1 \diff \lambda_2 .
\label{gtb}
\end{multline}
The recursive function $\phi$ is defined in~\eqref{Rec1}.  The
integral in~\eqref{gtb} can be evaluated by any numerical method.  We
remark that the expression for the average received power
in~\eqref{gtb} does not apply for $N_t = 2$ since the cdf derived in
Lemma~\ref{lemma_Nt} only applies when $N_t > 2$.  We find the optimal
feedback interval $K^*$ by maximizing the average received power
in~\eqref{1a2}, which is determined by \eqref{gtb}.  The same
conclusion made for the previous channel model on the maximum feedback
interval of the AFP scheme still applies for this channel model.
However, \cite{mimo} has shown that in order to maintain $\gamma_{N_t
  \times 2}$, $B$ needs to scale with $N_t$ as $N_t$ becomes large.
Otherwise, if $B/N_t \to 0$, then the quantization error of transmit
beamforming vector will be large and, hence the received power
$\gamma_{N_t \times 2}$ will be close to that with no CSI.  Thus, for
a fixed feedback rate, the maximum feedback interval of the AFP scheme
must increase as $N_t$ increases.

From the analysis, we see that optimizing the feedback interval
requires the temporal correlation coefficient $\alpha$, which in
practice, has to be estimated.  For instance, a least-square
estimator~\cite{zheng99} can be applied to determine $\alpha$.  Since
channel statistics does not change as often as channel realization
does, $\alpha$ may not need to be estimated frequently.

In this section, our analytical results only apply to channels with
either two transmit or two receiver antennas.  For channels with
arbitrary $N_t$ and $N_r$, the expression for the received power is
not tractable due to the pdf of $\bv^{\dag}\bL\bv$ and the joint pdf
of the ordered eigenvalues of $\bH(1)^\dag \bH(1)$.  However, the
performance of the system with an arbitrary number of antennas can be
well approximated by its performance in a large system regime to be
defined in the next section.

\section{Large System Analysis}
\label{large_sys}

The large system limit refers to one of which $N_t, N_r, B$ tend to
infinity with fixed $\Nr \triangleq N_r/N_t$ and $\B \triangleq
B/N_t$.  In a large system limit, the pdf of the ordered eigenvalues
converges to a deterministic function~\cite{tulino04} and hence,
performance analysis of systems with arbitrary size becomes
accessible.  It is shown by~\cite{mimo} that with some feedback ($\B >
0$) and fixed $\Nr$, the achievable rate defined in~\eqref{c_rvq}
increases with $\log(\rho N_t)$.  Thus, we define an achievable rate
difference as follows
\begin{align}
  R_{\triangle} & \triangleq R - \log(\rho N_t) \\ & = E \left[ \log
  \left(\frac{1}{\rho N_t} + \frac{1}{N_t} \bvh(k)^{\dag}
  \bH(k)^{\dag} \bH (k) \bvh (k) \right) \right] .
\end{align}
Therefore, $R_{\triangle}$ is a rate difference between an actual rate
and $\log(\rho N_t)$ and the difference increases with
$\B$~\cite{mimo}.  With feedback rate $\B$ per fading block, we apply
the AFP scheme described in Section~\ref{afp} and compute the average
rate difference over an interval of $K$ fading blocks given by
\begin{equation}
\bar{R}_{\triangle} = \frac{1}{K}\sum_{k=1}^{K}E\left[ \log
  \left(\frac{1}{\rho N_t} + \frac{1}{N_t}\bvh(1)^{\dag} \bH(k)^{\dag}
  \bH (k) \bvh (1) \right)\right]
\label{c_interval_large}
\end{equation}
where the quantized beamforming of the first block is used for the
whole interval of $K$ blocks.  We note that in the previous section,
we chose to evaluate the upper bound on the rate via the average
received power due to the intractability of the rate analysis.
However, in this section, we evaluate the rate difference.

\subsection{Large System With $\Nr > 0$}

First we consider the large system with $\Nr > 0$.  In other words,
the numbers of transmit and receive antennas are increasing at the same
rate.  Similar to the analysis of the system with a finite number of
antennas, we determine the received power per transmit antenna
$\frac{1}{N_t}\bvh^{\dag}(1) \bH^{\dag}(k) \bH (k) \bvh(1)$ by
applying the Gauss-Markov equation in~\eqref{eq_1od} and evaluate each
term after substitution.  The first of the two nonzero terms is shown
by~\cite{mimo,dai09} to converge in a large system limit
\begin{equation}
  \frac{1}{N_t} \bvh(1)^{\dag} \bH(1)^{\dag} \bH (1) \bvh
  (1) \longrightarrow \srvq\left(\B K\right)
\label{f1N}
\end{equation}
where $\B K$ is the normalized feedback bits used for quantizing
$\bvh(1)$ and the expression for the function $\srvq (x)$ is as
follows~\cite{mimo}.  Suppose
\begin{equation}
  \beta = \frac{1}{\log(2)} \left( \Nr \log
  \left(\frac{\sqrt{\Nr}}{1+\sqrt{\Nr}} \right) + \sqrt{\Nr} \right).
\label{bst}
\end{equation}
For $0 \le x \le \beta$, $\srvq$ satisfies
\begin{equation}
  \left( \srvq \right)^{\Nr} \me^{-\srvq} = 2^{-x} \left(
  \frac{\Nr}{\me} \right)^{\Nr}
\label{lsr}
\end{equation}
 and for $x \ge \beta$,
\begin{multline}
  \srvq(x) = (1 + \sqrt{\Nr})^2 - \exp \Big\{ \frac{1}{2} \Nr \log(\Nr) \\
   \quad - (\Nr-1) \log (1 + \sqrt{\Nr}) + \sqrt{\Nr} - x \log(2) \Big\} .
\label{lsr2}
\end{multline}
The second nonzero term can be shown to converge to
\begin{equation}
  \frac{1}{N_t} \bvh(1)^{\dag} \bW(k-i)^{\dag} \bW(k-i) \bvh(1)
  \longrightarrow \Nr .
\label{1Nv}
\end{equation}
Applying~\eqref{f1N} and~\eqref{1Nv}, we obtain
\begin{multline}
  \lim_{(N_t,N_r,B)\to \infty} \frac{1}{N_t} \bvh(1)^{\dag}
  \bH(k)^{\dag} \bH (k) \bvh(1)\\ = \Nr + \alpha^{2k-2} \left(
  \srvq\left(\B K \right) -\Nr \right). \label{1ap}
\end{multline}
Consequently, the expression for the asymptotic rate difference is given by
\begin{align}
  \Cbt^{\infty} &= \lim_{(N_t,N_r,B)\to \infty} \Cbt \\
               &= \frac{1}{K} \sum_{k=1}^K \log(\Nr + \alpha^{2k-2} \left(
  \srvq \left( \B K \right)-\Nr \right)) .
\label{acap}
\end{align}
We would like to maximize the asymptotic achievable rate difference
averaged over the feedback interval $K$.  For a given feedback rate of
$\B$ and $\Nr > 0$, the optimal feedback interval that maximizes the
asymptotic achievable rate difference is therefore given by
  \begin{equation}
      K^* = \arg \max_{K \in \mathbb{Z}^+} \left[ \prod_{k=1}^K \Nr +
        \alpha^{2(k-1)} \left( \srvq \left( \B K \right)-\Nr
        \right)\right]^{\frac{1}{K}} .
\label{Ksa}
  \end{equation}
Similar to a finite-size system, exhaustive search over some range of
$K$ can be used to obtain a suboptimal feedback interval.  We note
that the optimal feedback interval in \eqref{Ksa} will depend on the
temporal correlation coefficient, feedback rate, and the number of
transmit and receiver antennas.  Next we consider two extreme regimes
for which $\alpha \to 0$ and $\alpha \to 1$.  When the channel does
not change ($\alpha \to 1$), the optimal feedback interval $K^*$ can
be shown to be infinite from~\eqref{Ksa}.  This implies that only one
feedback update at the start with all available feedback bits giving
the maximum rate difference.

When the channel fades independently from a current block to the next
block ($\alpha \to 0$), the rate difference in~\eqref{acap} becomes
\begin{equation}
  \lim_{\alpha \to 0} \Cbt^{\infty} = \frac{1}{K} \log(\srvq \left( \B
  K \right)) +\frac{K-1}{K} \log(\Nr) . \label{acap1}
\end{equation}
Maximizing the rate-difference expression in~\eqref{acap1}, the
optimal feedback interval is given by
\begin{equation}
   K^* = \arg \max_{K\in \mathbb{Z}^+} \frac{1}{K} \log
   \left(\frac{\srvq (\B K )}{\Nr} \right) ,
\label{lm2}
\end{equation}
which depends on $\Nr$ and $\B$.  We remark that for moderate to large
$\B$, $K^* = 1$.  Hence, if the channel is temporally uncorrelated,
the feedback update must occur as frequent as possible.  In other
words, the MFP scheme will outperform the AFP scheme.

For general $\Nr$ and $\alpha$, to find the range of $K$ in which the
AFP scheme performs better than the MFP scheme, we solve for $K$
\begin{equation}
  \Cbt^{\infty} > \left. \Cbt^{\infty} \right\rvert_{K = 1} =
  \srvq(\B)
\end{equation}
where $\Cbt^{\infty}$ is stated in~\eqref{acap}.

\subsection{Large System With $\Nr \to 0$}

Next we examine the system in which $\Nr \to 0$ in a large system
limit.  The results will apply to the system in which the receiver is
equipped with only single antenna (MISO channel) or a fixed number of
antennas while the transmitter is equipped with much larger number of
antennas.  First we evaluate the large system limit of
$\frac{1}{N_t}\bvh(1)^{\dag} \bH(k)^{\dag} \bH (k) \bvh (1)$.  For
$\Nr = 0$, \cite{mimo} shows that
\begin{align}
  \frac{1}{N_t} \bvh(1)^{\dag} \bH(1)^{\dag} \bH (1) \bvh (1)
  &\longrightarrow \srvq\left(\B K\right)\\
  &= 1 - 2^{-\B K}
\end{align}
while
\begin{equation}
  \frac{1}{N_t} \bvh(1) ^{\dag}\bW(k-i)^{\dag} \bW(k-i) \bvh(1)
  \longrightarrow 0 .
\end{equation}
Thus, the asymptotic achievable rate difference is given by
\begin{align}
   \Cbt^{\infty} &= \frac{1}{K} \sum_{k=1}^K \log\left( \alpha^{2k-2}
   (1 - 2^{-\B K}) \right) \\ &= (K-1) \log (\alpha) + \log \left(1 -
   2^{-\B K}\right)
\label{cbn0}
\end{align}
for $0 < \alpha \le 1$.

Maximizing the asymptotic achievable rate difference in~\eqref{cbn0}
gives the optimal feedback interval as follows
\begin{equation}
    K^* = \arg \max_{K \in \mathbb{Z}^+}
    \ \alpha^{K-1} (1 - 2^{-\B K}) .
\label{th2}
\end{equation}
If the integer constraint is removed, we can find $K^*$ from the first
derivative of $\Cbt^{\infty}$ in~\eqref{cbn0} and obtain the following
approximation
\begin{equation}
  K^* \approx \frac{1}{\B} \log_2 \left(1 + \frac{\B \log 2}{ \log
    \frac{1}{\alpha}} \right)
\label{ksa}
\end{equation}
where $0 < \alpha < 1$.  The asymptotic $K^*$ obtained
from~\eqref{ksa} is close to that for a finite-size system.  We note
that for large feedback $\B$, $K^*$ is small.  The solution implies
that the feedback update should occur often when a large number of
feedback bits is available.  For a small-feedback regime ($\B \to 0$),
$K^*$ is approximated as follows
\begin{equation}
  \lim_{\B \to 0} K^* \approx \frac{\log 2}{\log \frac{1}{\alpha}} .
\end{equation}
We note that $K^*$ is increasing with $\alpha$.  Thus, we can conclude
that with a low feedback rate and a highly correlated channel, the
feedback interval should be large or the feedback update should occur
less frequently.

Comparing the rates obtained from the AFP and MFP schemes, we find
that the feedback interval $K$ for the AFP scheme must be larger than
\begin{equation}
  K > 1 + \frac{1}{\log \alpha}\log\left( \frac{1 - 2^{-\B}}{1 -
    2^{-\B K}}\right).
\end{equation}
Hence, as channels become less correlated (small $\alpha$), $K$ can be
large.  This bound is obtained by solving
\begin{equation}
  \Cbt^{\infty} > \left. \Cbt^{\infty} \right\rvert_{K = 1} =
  \log(1 - 2^{-B})
\end{equation}
where $\Cbt^{\infty}(K)$ is stated in~\eqref{cbn0}.

\section{Numerical Results}
\label{num_re}

\begin{figure}[h]
\centering
\includegraphics[width=3.7in]{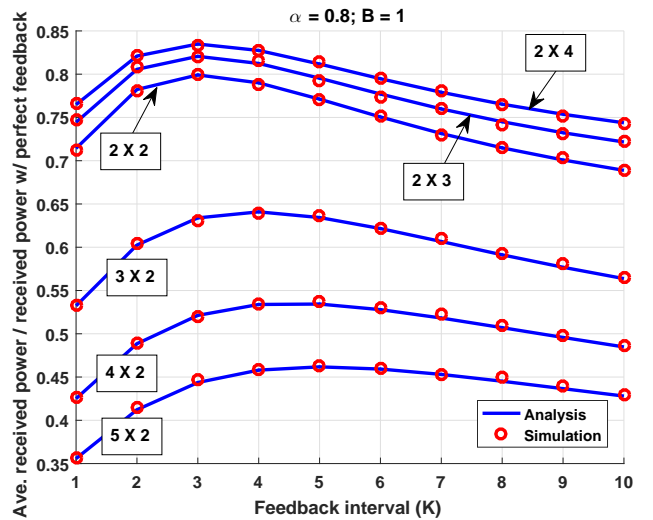}
\caption{The received power averaged over the feedback interval and
  normalized by the received power with infinite feedback, is plotted
  with the length of the feedback interval for various channels where
  $\alpha = 0.8$ and $B = 1$.  Both simulation and analytical results
  are shown.}
\label{OptNtNr}
\end{figure}

To illustrate the performance of the considered schemes, Monte Carlo
simulation is performed with 3,000 channel realizations.  First, we
compare the analytical results derived in Section~\ref{afp} with the
simulation results.  Fig.~\ref{OptNtNr} shows the average received
power normalized by the average received power with perfect feedback,
over the feedback interval of the AFP scheme with the feedback
interval $K$.  The feedback rate $B = 1$ bit per block and correlation
coefficient $\alpha = 0.8$.  We have two sets of system sizes.  For
the first set, $N_t$ is fixed at 2 with various $N_r$ ($2 \times 2$,
$2 \times 3$, and $2 \times 4$).  We see that the analytical result
in~\eqref{Ave2xNr}, which is shown with a solid line, perfectly
matches with the simulation result, which is shown with circles.  For
all $N_r$, the optimal feedback interval $K^*$ is 3.  Adding more
receive antennas will increase the received power since the receiver
can capture more transmitted signal.  With 4 receive antennas, the
system with $K^*$ achieves closer to 85\% of the performance with
infinite feedback.  The AFP scheme with the optimal $K$ ($K = 3$) can
outperform the MFP scheme ($K = 1$) by close to 11\%.

For the second set of system sizes in which $N_r = 2$ and $N_t$
varies ($2 \times 2$, $3 \times 2$, $4 \times 2$, and $5 \times 2$),
the analytical result comes from~\eqref{1a2}, and \eqref{gtb}.  We see
that the optimal interval $K^*$ increases with $N_t$ since the number
of bits ($BK$) required to quantize the beamforming vector increases
with $N_t$.  For a larger system ($5 \times 2$), the AFP with $K^* =5$
can outperform the MFP by as much as 27\%.

For $2 \times 2$ channels, we see that the AFP scheme with $2 \le K
\le 7$ gives larger averaged received power than the MFP scheme. The
range of $K$ is accurately predicted by~\eqref{Klf}.  For the $2
\times 4$ channel, the range of $K$ for which the AFP performs better
is $2 \le K \le 8$, which can also be obtained by~\eqref{Klf}.

\begin{figure}[h]
\centering
\includegraphics[width=3.65in]{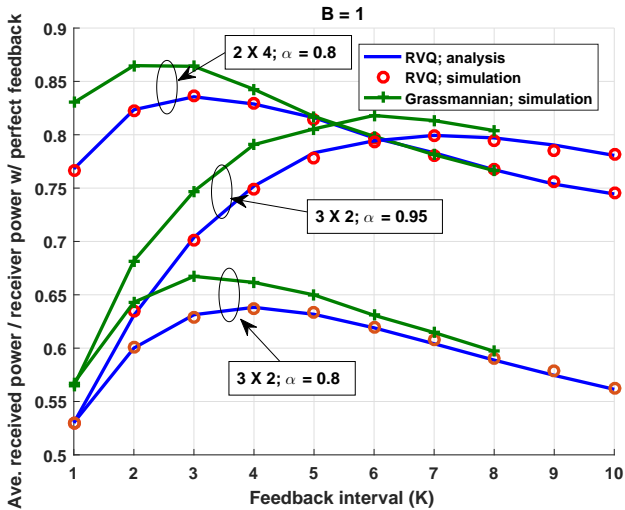}
\caption{Averaged received power normalized by the received power with
  the perfect feedback is shown for both RVQ and Grassmannian
  codebooks with varying $K$.}
\label{OptNtNrGrass}
\end{figure}

In Fig.~\ref{OptNtNrGrass}, we compare the performance of RVQ codebook
with that of the Grassmannian codebook~\cite{love03}, which maximizes
the minimum chordal distance between any two codebook entries.  The
Grassmannian codebook is optimal for channels with finite number of
antennas and hence, is shown in the figure to outperform RVQ codebook.
However, the Grassmannian codebook is more complex to construct than
RVQ codebook especially when the number of entries is large.  Thus, in
the figure, we do not have results of the Grassmannian codebook beyond
$K = 8$.  We note that the performance shown in
Fig.~\ref{OptNtNrGrass} is the averaged received power normalized by
the received power with infinite feedback.  We see a larger
performance gap between the two codebooks when $BK$ is small or when
the number of quantization bits is small.  For all 3 cases shown,
$K^*$ for RVQ codebook and the optimal $K$ that maximizes the received
power for Grassmannian codebook only differs by 1.  This implies that
the optimal feedback interval derived for RVQ codebook in this study
can be applied to the Grassmannian codebook with some small
degradation.  For the $3 \times 2$ channel, we see that the gain of
AFP with the optimal $K$ over MFP ($K = 1$) increases when the channel
becomes more correlated ($\alpha$ closer to 1). For the 3 x 2 channel
with $\alpha = 0.95$, the Grassmannian codebook with $K^* = 7$ ($K^*$
is derived with RVQ codebook) achieves approximately 82\% of the rate
with perfect feedback while the Grassmannian codebook with $K = 1$ or
the MFP scheme achieves only 57\% of the rate with perfect
feedback. Thus, the performance gain of the AFP scheme over the MFP
scheme in this instance is about 44\%.

\begin{figure}[h]
\centering
\includegraphics[width=3.65in]{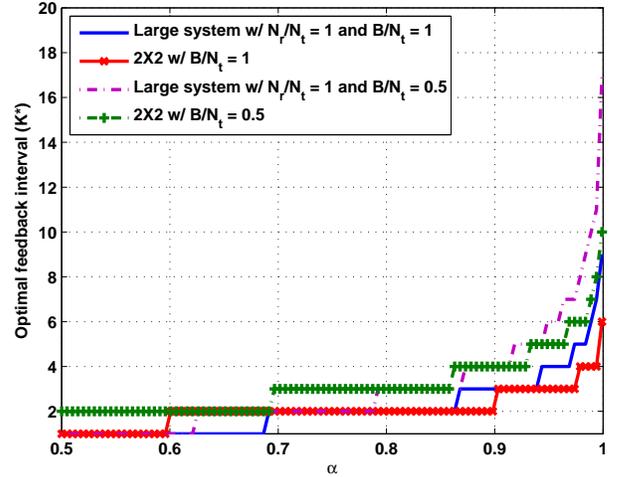}
\caption{Optimal feedback intervals for a large system with $\Nr = 1$
  obtained by~\eqref{Ksa} and for a $2\times 2$ channel obtained by
  maximizing \eqref{Ave2xNr} are shown with varying channel correlation
  $\alpha$ and feedback rate.}
\label{OptLargeFinite}
\end{figure}

Fig.~\ref{OptLargeFinite} shows the optimal feedback interval $K^*$
for a $2 \times 2$ channel with different values of correlation
coefficient $\alpha$ and the number of feedback bits per fading block
per transmit antenna $B/N_t$.  We consider a mobile system operating
at 900 MHz with 5-ms average fading block for which $\alpha$ varies
from 0.5 to 0.9999 as the speed of mobile decreases from 60 km/h to 1
km/h~\cite{jake93}.  We see that for a slow fading channel $(\alpha
\rightarrow 1)$, feedback update can be less frequent and thus, the
feedback interval is large.  On the other hand, fast fading channels
(smaller $\alpha$) require frequent feedback updates.  If the feedback
rate per transmit antenna ($B/N_t$) is increased from 0.5 to 1, we see
that $K^*$ decreases.

In Fig.~\ref{OptLargeFinite}, we also show the optimal interval $K^*$
of a large system with $\Nr = 1$ obtained by solving~\eqref{Ksa}.  We
remark that $K^*$ for a large system is obtained by maximizing the
rate difference while $K^*$ for a $2 \times 2$ channel is obtained by
maximizing the averaged received power.  However, we see that the
large system results can give a good approximation of those of a very
small system.

\begin{figure}[h]
\centering
\includegraphics[width=3.65in]{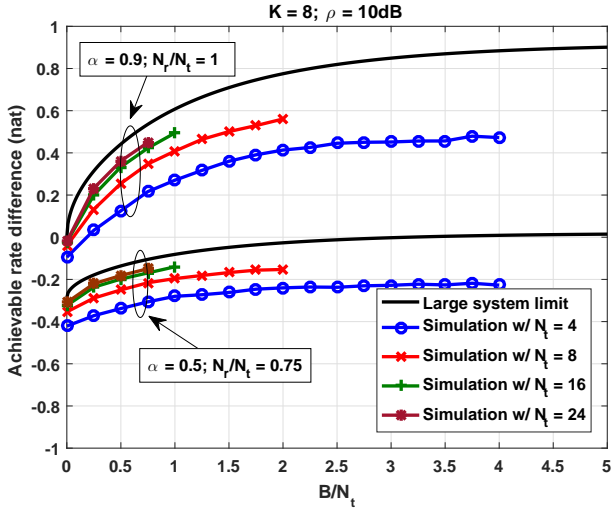}
\caption{Averaged rate difference for a large system is compared with
  that for a finite-size system with $K=8$ and $\rho = 10$ dB.}
\label{LargSysBit}
\end{figure}

In Fig.~\ref{LargSysBit}, we compare the achievable rate difference of
a large system derived in~\eqref{acap} with that of a finite-size
system for various feedback rates per transmit antenna $\B$.  The
feedback interval $K$ is fixed at 8 blocks and SNR $\rho$ is at 10 dB.
The averaged rate gain of finite-size and large systems is obtained
from~\eqref{c_interval_large} and~\eqref{acap}, respectively.  We see
that as the system size increases from $N_t = 4$ to 8, 16, and 24, the
simulation results approach the large system results.  However, we
note that the convergence to the asymptotic results is slow.  Thus,
unless the system size is very large, the gap between the actual and
the asymptotic rate difference might be significant.  We also note
that the rate difference increases with $\B$ as expected, but rate of
increase is different for different values of $\alpha$.  When the channel is
less correlated ($\alpha = 0.5$), the quantized beamforming vector of
the first block is not a good substitute for that of the next blocks.
Consequently, we do not see much increase in that case although the
feedback rate is increased.  On the contrary, we see a large increase
when the channel is more correlated ($\alpha = 0.9$) since the
quantized beamforming vector of the first block performs well for all
subsequent blocks in the same interval.  Since we quantize beamforming
vectors with the RVQ codebook, which requires an exhaustive search to
find the quantized vector, the search complexity can be too large for
large $B$.  Thus, some of the plots in Fig.~\ref{LargSysBit} do not
extend to a larger feedback rate.

\begin{figure}[h]
\centering
\includegraphics[width=3.65in]{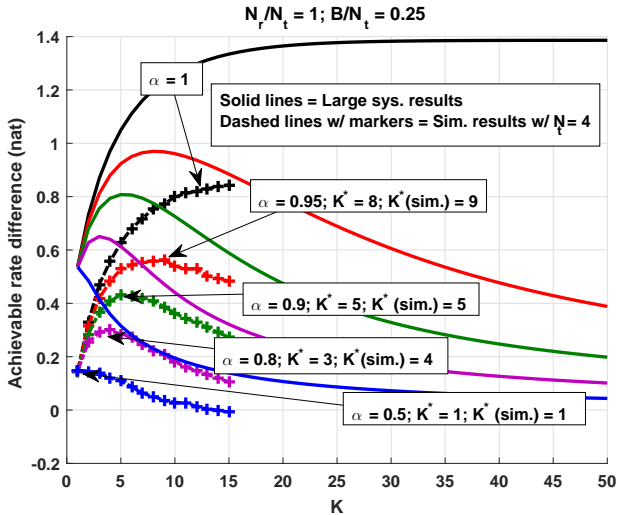}
\caption{Averaged rate difference for a large system with $\Nr = 1$ is
  compared with that for a $4 \times 4$ channel with $\B=0.25$ and
  $\rho = 10$ dB.}
\label{mimo_max}
\end{figure}

\begin{figure}[h]
\centering
\includegraphics[width=3.65in]{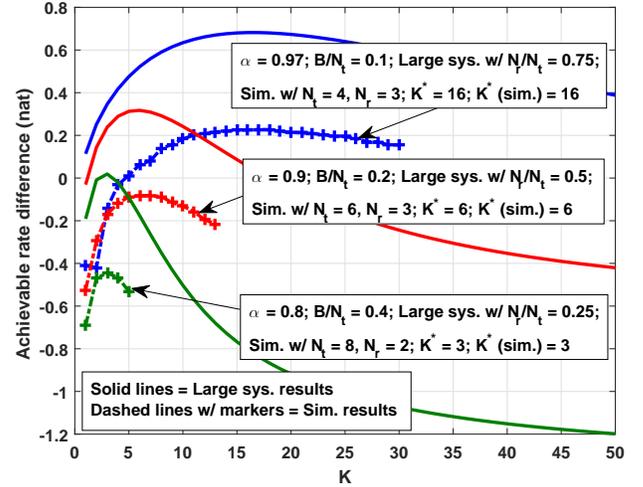}
\caption{Averaged rate difference for a large system is compared with
  systems with different sizes, feedback rates, and $\alpha$.}
\label{mimo_max2}
\end{figure}

In Fig.~\ref{mimo_max}, we set $\B = 0.25$ and vary $K$ for channels
with different temporal correlation.  We compare the rate difference
of $4 \times 4$ channels and that of a large system with $\Nr = 1$.
For a $4 \times 4$ channel with $\alpha = 0.9$, the AFP scheme with $K
= 5$ performs almost twice as much as the MFP scheme does (the green
line with pluses).  For time-invariant channels ($\alpha = 1$), the
optimal $K$ is large.  Although the difference between the results of
small-size and large systems can be large as shown in
Fig.~\ref{LargSysBit}, the optimal feedback interval $K^*$ obtained
from the two results is close (either off by 1 or identical).  We also
compare the optimal $K$ from the simulation and analytical results
with different system sizes, feedback rates, and channel correlation
coefficients in Fig.~\ref{mimo_max2}.  The results reinforce that the
optimal feedback interval that maximizes the rate difference of a
finite-size system, can be predicted quite accurately by the
large-system analysis.

\begin{figure}[h]
\centering
\includegraphics[width=3.65in]{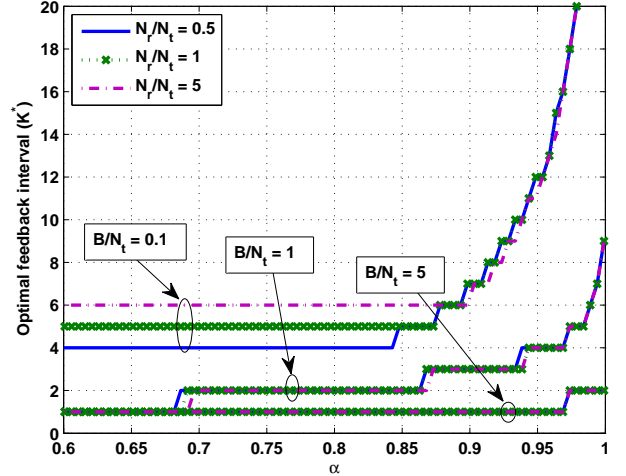}
\caption{The optimal feedback interval is shown with $\alpha$ and $\B$
  for large-system channels with different $\Nr$.}
\label{mimo_nr}
\end{figure}

We plot the optimal feedback interval $K^*$ with the temporal
correlation $\alpha$ for large-system channels with different $\Nr$
and $\B$ in Fig.~\ref{mimo_nr}.  The same trend as shown in
Fig.~\ref{OptLargeFinite} is also observed in this figure. $K^*$
increases with $\alpha$.  However, we note that $K^*$ is mostly
unchanged across different values of $\Nr$, except when $\B$ is
extremely low.  Similar observation regarding to different number of
receive antennas was also noted for a finite-size system.

\begin{figure}[h]
\centering
\includegraphics[width=3.65in]{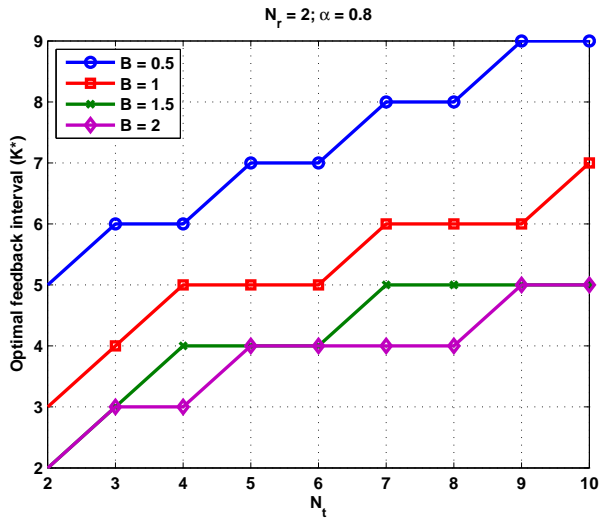}
\caption{Optimal feedback intervals $K^*$ for various channel
  sizes (with fixed $N_r = 2$) are plotted with the number of transmit
  antennas $N_t$ and the number of feedback bits per fading block
  $B$.}
\label{OptVsNt}
\end{figure}

Fig.~\ref{OptVsNt} shows how the optimal feedback interval $K^*$
increases with the number of transmit antennas $N_t$, but decreases
with the number of feedback bits per fading block $B$. We note that
$K^*$ is obtained by first substituting~\eqref{1a2} into~\eqref{optK1}
and then solving~\eqref{optK1} numerically.  We set $N_r = 2$ and
$\alpha = 0.8$.  For larger $N_t$, the number of bits to quantize the
beamforming vector needs to increase to maintain the rate performance
and hence, the feedback interval has to increase as well.  Similar to
the results in Fig.~\ref{OptLargeFinite}, as $B$ increases, the
feedback interval can be reduced.

\begin{figure}[h]
\centering
\includegraphics[width=3.65in]{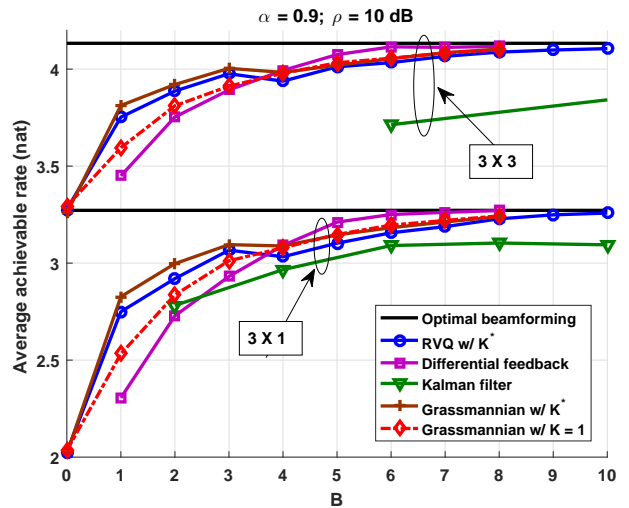}
\caption{Average rate of various feedback schemes for $3 \times 1$ and
  $3 \times 3$ channels are shown with the number of feedback bits per
  fading block.  $\alpha=0.9$ and SNR = 10 dB.}
\label{CompareAllFinal}
\end{figure}

In Fig.~\ref{CompareAllFinal}, we compare the AFP and MFP schemes with
existing Kalman-filter scheme and differential-feedback scheme in the
literature for $3 \times 1$ and $3 \times 3$ channels.
In~\cite{mehanna14}, KF scheme is applied to construct the channel
vector (or channel matrix) at the transmitter, which then can compute
the optimal transmit beamforming.  For a fair comparison, we assume
that the channel estimation at the receiver is perfect.  The receiver
quantizes an innovation term, which is the difference between the
received signal and its estimate based on channel estimates from the
previous blocks.  The innovation can be straightforwardly shown to be
zero-mean Gaussian with some finite variance.  Thus, for
quantization, we apply a generalized Lloyd algorithm~\cite{lloyd82},
which minimizes the mean square error.  The quantized innovation is
fed back to the transmitter for every fading block.  To construct the
channel vector, we follow the steps in~\cite{mehanna14,xu14}.  The
performance of KF scheme is shown in Fig.~\ref{CompareAllFinal}.

For the performance of differential feedback, we apply method 1
in~\cite{kim11dif}.  The codebook that quantizes transmit beamforming
vector is not fixed, but is gradually updated by the rotation matrix
selected from a rotation codebook and the normalized radius, which is
a function of $N_t$, $B$, $\alpha$, and block index $k$. The rotation
codebook consists of unitary matrices.  For the optimal rotation
codebook, the minimum distance defined by~\cite{kim11dif} between two
entries is maximized.  For the results in this figure, we generate
10000 random codebooks with the desired number of entries and find the
codebook with the largest minimum distance between any two codebook
entries. Thus, our rotation codebook is suboptimal, but should be
close to the optimum due to a larger number of trials.

In Fig.~\ref{CompareAllFinal}, we $\alpha = 0.9$ and SNR = 10 dB.  We
note that some feedback schemes may require some initial feedback bits
and thus, their performance does not extend to $B = 0$ or small $B$.
For example, the KF scheme needs at least $2N_r$ bits to quantize an
innovation, which is an $N_r$-dimensional complex vector.  From the
figure, we see that, Grassmannian codebook with $K^*$, which is
obtained from our analysis, performs the best for low to moderate
feedback rates and is followed closely by RVQ codebook with $K^*$.
For the $3 \times 1$ channel with $B=2$, the Grassmannian codebook
with $K^*$ outperforms KF scheme by about 10\%.  The differential
feedback scheme by~\cite{kim11dif} performs better than other schemes
when feedback rate is larger and performs worse when feedback rate is
small.  As mentioned in~\cite{kim11dif}, the scheme requires some
sufficient feedback to compensate for cumulative quantization error.
We see that codebooks with $K^*$ outperform the KF scheme for all
feedback rates for both $3 \times 1$ and $3 \times 3$ channels.
Performance degradation is quite significant for the KF scheme when
applied to MIMO channels. If feedback is not sufficient, the KF scheme
does not track channel matrix well and hence, produces transmit
beamforming, which is not aligned with the strongest channel mode.

\section{Conclusions}
\label{conclude}

We have analyzed the feedback interval that maximizes either the average
received power or the rate difference for MIMO channels.  For the channel model
with either two transmit or two receive antennas, the optimal interval
depends more on the channel correlation, the number of transmit
antennas, and the feedback rate, and less on the number of receive
antennas.  For that model, we formulated the received-power maximizing
problem in which the exact feedback interval can be found.  For
systems with arbitrary number of transmit and receive antennas, large
system analysis can be used to predict the optimal interval accurately
as shown by the numerical examples.  The optimal feedback interval is
a function of the channel correlation, the number of feedback bits per
antenna, and the ratio between the number of transmit and receive
antennas.  However, the optimal feedback interval also is less
sensitive to the change in the number of receive antennas.

When the feedback rate is low, the AFP scheme with the optimal
feedback interval outperforms the other schemes including the KF
scheme and differential feedback scheme.  The performance gain of the
AFP scheme over the other schemes can be as much as 10\%.  Thus, the
feedback interval should be adapted according to channel condition.
However, when the feedback rate is high, the performance difference
among the different schemes may not be significant.  We also note that
the optimal feedback interval derived for RVQ codebook and be applied
with Grassmannian codebook, which is optimal for finite-size channels,
with small degradation.

In this work, we assume that training of the channel is sufficient and
thus, CSI at the receiver is perfect.  For a system with limited
training, the actual performance of the AFP scheme will be lower than
that obtained in the paper and the KF scheme may perform better.
Since we only consider a point-to-point channel in the present work,
broadcast or multiple-access channels are also of interest and can be
considered in future work.

\appendix

\subsection{Proof of Lemma~\ref{lemma_Nr}}
\label{a1}

Apply the Gauss-Markov model in~\eqref{eq_1od} and some algebraic
manipulation to obtain
\begin{multline}
  E\left[ \bvh(1)^{\dag} \bH(k)^{\dag} \bH (k) \bvh (1)\right]\\ =
  \alpha^{2k-2} E\left[ \bvh(1)^{\dag} \bH(1)^{\dag} \bH (1)
    \bvh(1) \right] \\ + (1-\alpha^2) \sum_{i=0}^{k-2} \alpha^{2i}
  E\left[ \bvh(1)^{\dag} \bW(k-i)^{\dag} \bW (k-i) \bvh (1)\right]
\label{eq_o}
\end{multline}
where the expectation of the cross term that consists of $\bW$ is
equal to zero since $\bW$ has zero mean and is independent of $\bvh
(1)$ and all channel matrices $\bH (k), \forall k$.

We proceed to analyze the first expectation in~\eqref{eq_o}.
Following the same argument pertaining to the received power
in~\eqref{maxv}, $\bv^{\dag}_i \bH^{\dag}(1) \bH(1) \bv_i$ is
independent and has the same distribution as $\bv^{\dag}_i \bL
\bv_i$~\cite{veeravalli13} where $\bL =
\text{diag}([\lambda_1,\lambda_2])$, and $\lambda_1$ and $\lambda_2$
are the ordered eigenvalues of $\bH(1)^{\dag} \bH(1)$ with $\lambda_1
\ge \lambda_2 \ge 0$.  The cdf for $\bv^{\dag}_i \bL \bv_i$
conditioned on $\lambda_1$ and $\lambda_2$ is given
by~\cite{veeravalli13}
\begin{equation}
  F_{\bv^{\dag}_i \bL \bv_i | \lambda_1, \lambda_2} (x) =
  \left\{ \begin{array}{l@{\quad:\quad}l}
   0 & 0 \le x < \lambda_2\\
    \frac{x-\lambda_2}{\lambda_1-\lambda_2} & \lambda_2 \leq x \leq
    \lambda_1\\
   1 & x > \lambda_1 \end{array} \right. .
\end{equation}
We then apply integration by parts to obtain the expression for the
conditional expectation as follows
\begin{align}
  E &\left[ \bvh(1)^{\dag} \bH(1)^{\dag} \bH(1) \bvh (1) \mid \lambda_1,
    \lambda_2 \right] \nonumber\\
  &= \lambda_1 - \int_{\lambda_2}^{\lambda_1}
  \left( F_{\bv^{\dag}_i \bL \bv_i | \lambda_1, \lambda_2}(x)
  \right)^{2^{BK}} \, \diff x\\
  &= \lambda_1 - \frac{\lambda_1 - \lambda_2}{2^{BK} + 1}.
\end{align}

Averaging over the two eigenvalues gives
\begin{multline}
  E \left[ \bvh(1)^{\dag} \bH(1)^{\dag} \bH(1) \bvh(1) \right]\\
  =
  \int_0^\infty \int_0^{\lambda_1} \left(\lambda_1 - \frac{\lambda_1 -
    \lambda_2}{2^{BK} + 1}\right) f_{\bL}(\lambda_1,\lambda_2) \,
  \diff \lambda_2 \diff \lambda_1 \label{Ef}
\end{multline}
where $f_{\bL}(\lambda_1,\lambda_2)$ is a joint pdf for the two
ordered eigenvalues of a Wishart matrix $\bH(1)^{\dag} \bH (1)$ given
by~\cite{fisher39}
\begin{multline}
   f_{\bL}(\lambda_1,\lambda_2) = \frac{\lambda_1^{N_r-2}
     \lambda_2^{N_r-2} (\lambda_1-\lambda_2)^2 \me^{-(\lambda_1 +
       \lambda_2)}}{(N_r-1)!(N_r-2)!},\\ \lambda_1 \ge \lambda_2
   \ge 0.
\label{pdln}
\end{multline}
Suppose
\begin{equation}
  \phi(m,n) \triangleq \int_{0}^{\infty} \lambda_1^{m}
  \me^{-\lambda_1} \int_{0}^{\lambda_1} \lambda_2^{n} \me^{-\lambda_2}
  \, \diff \lambda_2 \diff \lambda_1, \quad \forall m,n \ge 1.
\label{pmn}
\end{equation}
By substituting~\eqref{pdln} into~\eqref{Ef} and rearranging the terms,
we can write $E \left[ \bvh(1)^{\dag} \bH(1)^{\dag} \bH(1) \bvh(1)
  \right]$ in~\eqref{Ef} in terms of $\phi(\cdot, \cdot)$ as shown
in~\eqref{GNr}.

To evaluate~\eqref{pmn}, we apply integration by parts to the inner
integral to obtain
\begin{align}
  \phi(m,n) &= n \int_{0}^{\infty} \lambda_1^m \me^{-\lambda_1}
  \int_{0}^{\lambda_1} \lambda_2^{n-1} \me^{-\lambda_2} \, \diff
  \lambda_2 \diff \lambda_1 \nonumber\\
  &\quad - \int_{0}^{\infty}
  \lambda_1^{m+n}\me^{-2\lambda_1} \, \diff
  \lambda_1 \label{inner1}\\
   & =n \int_{0}^{\infty} \lambda_2^{n-1}
  \me^{-\lambda_2} \int_{\lambda_2}^{\infty} \lambda_1^m
  \me^{-\lambda_1} \, \diff \lambda_1 \diff \lambda_2 \nonumber \\
  &\quad - \frac{(m +
    n)!}{2^{m+n+1}} \label{inner}
\end{align}
where in~\eqref{inner1}, we switch the order of integration for the
first integral and evaluate the second integral.  We again evaluate
the inner integral in~\eqref{inner} and switch the order of
integration to obtain
\begin{multline}
  \phi(m,n) = m n \underbrace{\int_{0}^{\infty} \lambda_1^{m-1}
    \me^{-\lambda_1} \int_{0}^{\lambda_1} \lambda_2^{n-1}
    \me^{-\lambda_2} \, \diff \lambda_2 \diff
    \lambda_1}_{\phi(m-1,n-1)}\\ + \frac{m(m + n -1)!}{2^{m+n}} -
  \frac{(m + n)!}{2^{m+n+1}} .
\label{fmn}
\end{multline}
Adding the last two terms in~\eqref{fmn} gives \eqref{Rec1}.  The
initial conditions are obtained by evaluating the double integral
in~\eqref{pmn}.

Since $\bW(k-i)$ and $\bvh(1)$ are independent and
$E [\bW(k-i)^{\dag} \bW(k-i)]= N_r\bI$, we have that
\begin{align}
  E&\left[ \bvh(1)^{\dag} \bW(k-i)^{\dag} \bW(k-i) \bvh(1) \right]\nonumber\\
  &= E\left[ \Tr\{\bW(k-i)^{\dag} \bW(k-i)\bvh(1)\bvh(1)^{\dag}\} \right]\\
  &=\Tr\{ E[\bW(k-i)^{\dag} \bW(k-i)] E[\bvh(1)\bvh(1)^{\dag}]\}\\
  &= N_r \Tr\{E[\bvh(1)\bvh(1)^{\dag}]\}\\
  &= N_r \label{bwvn}
\end{align}
where $\Tr\{\cdot\}$ denotes the trace operator.

Finally, we substitute~\eqref{fmn} and~\eqref{bwvn} into~\eqref{eq_o}
and simplify to obtain \eqref{GkNr}.

\subsection{Proof of Lemma~\ref{lemma_Nt}}
\label{a2}

Since the considered matrix $\bH(1)^{\dag}\bH(1)$ has rank 2,
\begin{equation}
  \lambda_3 = \lambda_4 = \cdots = \lambda_{N_t} = 0.
\end{equation}
Applying the results from \cite[eq. (18)]{veeravalli13}, we obtain in
Lemma~\ref{lemma_Nt} the expression for the cdf $F_{\bv^{\dag} \bL
  \bv}(x)$ for $\lambda_2 \le x \le \lambda_1$ only.  Next we derive
the expression of the cdf when $0\leq x \leq \lambda_2$.  The
derivation is inspired by~\cite{mukkavilli03} where evaluating the cdf
$F_{\bv^{\dag}\bL\bv} (x)$ was formulated as finding the surface area
of an $N_t$-dimensional spherical cap.  The results
in~\cite{mukkavilli03} apply when $\bL$ is full rank.  In our case,
$\bL$ has rank 2 with nonzero diagonal entries $\lambda_1$ and
$\lambda_2$.

Recall that $\bv = [v_1 \ v_2 \ \cdots \ v_{N_t}]^T$ is an $N_t \times
1$ isotropically distributed vector with unit norm. Therefore, we have
\begin{multline}
  \Pr \{\bv^{\dag}\bL\ \bv \ge x\} \\
  = \Pr \left\{ \lambda_1 |v_1|^2 +
  \lambda_2 |v_2|^2 \ge x, \sum_{i = 1}^{N_t} |v_i|^2 = 1 \right\} .
\label{bvl}
\end{multline}
In $N_t$-dimensional space, we can view the set of vectors $\{ \bv \in
\mathbb{C}^{N_t} | \sum_{i = 1}^{N_t} |v_i|^2 = 1\}$ as a surface of
an $N_t$-dimensional unit ball centered at the origin.  We can rearrange
$\lambda_1 |v_1|^2 + \lambda_2 |v_2|^2 \ge x$ as follows
\begin{equation}
  \frac{|v_1|^2}{\frac{x}{\lambda_1}} +
  \frac{|v_2|^2}{\frac{x}{\lambda_2}} \ge 1 .
\end{equation}
The above inequality describes the region outside a two-dimensional
ellipse centered at the origin.  Because $\lambda_1 \ge \lambda_2$, the
widest part of the ellipse is determined by $\frac{x}{\lambda_2}$.
Since we consider the regime where $0 \le x \le \lambda_2$ or $0 \le
\frac{x}{\lambda_2} \le 1$, geometrically, the ellipse is completely
contained in the $N_t$-dimensional unit ball.

We take the same analytical approach as the one in~\cite{mukkavilli03}
by first finding the volume of the $N_t$-dimensional object prescribed by
$\lambda_1 |v_1|^2 + \lambda_2 |v_2|^2 \ge x$ and $\sum_{i = 1}^{N_t}
|v_i|^2 \le r^2$ where $r \ge 1$. (In the final steps, we will set $r
= 1$.)  Then, we compute its surface area, which is shown to be
proportional to the desired cdf $F_{\bv^{\dag}\bL\bv} (x)$
\cite{mukkavilli03}.

The volume of the region $\{\bv \in \mathbb{C}^{N_t} | \lambda_1
|v_1|^2 + \lambda_2 |v_2|^2 \ge x, \|\bv\|^2 \le r^2\}$ is denoted by
\begin{multline}
  \textsf{Vol}(\lambda_1 |v_1|^2 + \lambda_2 |v_2|^2 \ge x, \|\bv\|^2
  \le r^2) = \textsf{Vol}(\|\bv\|^2 \le r^2) \\- \textsf{Vol}(\lambda_1
  |v_1|^2 + \lambda_2 |v_2|^2 \le x, \|\bv\|^2 \le r^2)
\end{multline}
where the volume of an $N_t$-dimensional ball with radius $r$ is given
by~\cite{mukkavilli03}
\begin{equation}
   \textsf{Vol}(\|\bv\|^2 \le r^2) = \frac{\pi^{N_t}
     r^{2N_t}}{\Gamma(N_t+1)}
\label{vol}
\end{equation}
and $\textsf{Vol}(\lambda_1 |v_1|^2 + \lambda_2 |v_2|^2 \le x,
\|\bv\|^2 \le r^2)$ is the volume of the ellipsoid that is completely
contained in the hyperball with radius $r$.

To compute the volume of the ellipsoid, we first apply the following
transformation
\begin{equation}
  v_i = r_i \me^{j\theta_i}, \quad \forall i
\end{equation}
where $r_i$ and $\theta_i$ are the magnitude and phase of $v_i$,
respectively. Using spherical coordinates, the volume of the ellipsoid
is given by
\begin{multline}
   \textsf{Vol}(\lambda_1 |v_1|^2 + \lambda_2 |v_2|^2 \le x, \|\bv\|^2
   \le r^2)\\ = (2\pi)^{N_t} \int_{r_1 =
     0}^{\sqrt{\frac{x}{\lambda_1}}} \int_{r_2 = 0}^{\sqrt{\frac{x -
         \lambda_1r_1^2}{\lambda_2}}}\\ \left( \idotsint\limits_{\sum_{n
       = 3}^{N_t} r_n^2 \le r^2 - r_1^2 - r_2^2} r_3 \cdots r_{N_t} \,
   \diff r_3 \cdots \diff r_{N_t} \right) r_1 r_2 \, \diff r_2 \diff r_1 .
\label{vaa}
\end{multline}
We note that the multiple integral in the brackets in~\eqref{vaa} is
the volume of an $(N_t-2)$-dimensional ball with radius $r^2 - r_1^2 -
r_2^2$.  Applying~\eqref{vol}, we have
\begin{align}
&\textsf{Vol}(\bv^{\dag}\bL\ \bv \leq x,\|\bv\|^2 \leq r^2)
\nonumber\\ &= \frac{(2\pi)^2\pi^{N_t-2}}{\Gamma(N_t-1)} \int_{r_1 =
  0}^{\sqrt{\frac{x}{\lambda_1}}} \int_{r_2 =
  0}^{\sqrt{\frac{x-\lambda_1 r_1^2}{\lambda_2}}} r_1 r_2 \nonumber\\
&\quad \times (r^2 - r_1^2
-r_2^2) \, \diff r_2 \diff r_1
\\ &=\frac{\lambda_2}{2N_t(\lambda_1-\lambda_2)} \left( \left( r^2 -
\frac{x}{\lambda_1}\right)^{N_t} - \left( r^2-\frac{x}{\lambda_2}
\right)^{N_t} \right).\label{vol3}
\end{align}

To compute the surface area of the volume, we differentiate the volume
as follows
\begin{align}
  \textsf{Area}&(\bv^{\dag}\bL\ \bv \leq x,\|\bv\|^2 \leq 1) \nonumber\\
  &=
\left. \frac{\partial }{\partial x \partial r^2}
\textsf{Vol}(\bv^{\dag}\bL\ \bv \leq x,\|\bv\|^2 \leq
r^2)\right|_{r^2=1} \\ &=
\frac{\pi^{N_t}}{(N_t-2)!(\lambda_1-\lambda_2)}\nonumber\\
&\quad \times \left(
\left(1-\frac{x}{\lambda_2}\right)^{N_t-2}-
\left(1-\frac{x}{\lambda_1}\right)^{N_t-2} \right).
\end{align}
The surface area of the $N_t$-dimensional unit ball is given by
\begin{align}
\textsf{Area}(\|\bv\|^2 \le 1) &= \left. \frac{\partial}{\partial
  r^2}\textsf{Vol}(\|\bv\|^2 \le r^2)\right|_{r^2 = 1}\\
  &=\frac{\pi^{N_t}N_t}{\Gamma(N_t+1)}.\label{ArSp}
\end{align}

The pdf of $\bv^{\dag} \bL \bv$ is given by~\cite[eq. (115)]{veeravalli13}
\begin{align}
f_{\bv^{\dag} \bL \bv}(x) &= -\frac{\textsf{Area}(\bv^{\dag}\bL\ \bv
  \leq x,\|\bv\|^2 \leq 1)}{\textsf{Area}(\|\bv\|^2 \le 1)}
\\ &=\frac{N_t-1}{\lambda_1-\lambda_2}\left(
\left(1-\frac{x}{\lambda_1}\right)^{N_t-2} -
\left(1-\frac{x}{\lambda_2}\right)^{N_t-2} \right).\label{pdfNt}
\end{align}
Finally, the expression of the cdf for $0 \le x \le \lambda_2$
in~\eqref{cdfNt} can be obtained by integrating the pdf
in~\eqref{pdfNt}.

\bibliographystyle{IEEEtran}
\bibliography{IEEEabrv,ref}

\begin{thebibliography}{10}
\providecommand{\url}[1]{#1}
\csname url@samestyle\endcsname
\providecommand{\newblock}{\relax}
\providecommand{\bibinfo}[2]{#2}
\providecommand{\BIBentrySTDinterwordspacing}{\spaceskip=0pt\relax}
\providecommand{\BIBentryALTinterwordstretchfactor}{4}
\providecommand{\BIBentryALTinterwordspacing}{\spaceskip=\fontdimen2\font plus
\BIBentryALTinterwordstretchfactor\fontdimen3\font minus
  \fontdimen4\font\relax}
\providecommand{\BIBforeignlanguage}[2]{{%
\expandafter\ifx\csname l@#1\endcsname\relax
\typeout{** WARNING: IEEEtran.bst: No hyphenation pattern has been}%
\typeout{** loaded for the language `#1'. Using the pattern for}%
\typeout{** the default language instead.}%
\else
\language=\csname l@#1\endcsname
\fi
#2}}
\providecommand{\BIBdecl}{\relax}
\BIBdecl

\bibitem{gc2011}
W.~Santipach and K.~Mamat, ``Optimal feedback interval for
  temporally-correlated multiantenna channel,'' in \emph{Proc. IEEE Global
  Telecommun. Conf. (GLOBECOM)}, Houston, Texas, USA, Dec 2011, pp. 1--5.

\bibitem{telatar}
{\.I}.~E. Telatar, ``Capacity of multi-antenna {G}aussian channels,''
  \emph{European Trans. on Telecommun.}, vol.~10, pp. 585--595, Nov. 1999.

\bibitem{foschini98}
G.~J. Foschini and M.~J. Gans, ``On limits of wireless communications in a
  fading environment when using multiple antennas,'' \emph{Wireless Personal
  Commun.}, vol.~6, no.~3, pp. 311--335, Mar. 1998.

\bibitem{love08}
D.~J. Love, R.~W. Heath, Jr., V.~K.~N. Lau, D.~Gesbert, B.~D. Rao, and
  M.~Andrews, ``An overview of limited feedback wireless communication
  systems,'' \emph{{IEEE} J. Sel. Areas Commun.}, vol.~26, no.~8, pp.
  1341--1365, Oct. 2008.

\bibitem{love03}
D.~J. Love, R.~W. Heath, Jr., and T.~Strohmer, ``Grassmannian beamforming for
  multiple-input multiple-output wireless systems,'' \emph{{IEEE} Trans. Inf.
  Theory}, vol.~49, no.~10, pp. 2735--2747, Oct. 2003.

\bibitem{mimo}
W.~Santipach and M.~L. Honig, ``Capacity of a multiple-antenna fading channel
  with a quantized precoding matrix,'' \emph{{IEEE} Trans. Inf. Theory},
  vol.~55, no.~3, pp. 1218--1234, Mar. 2009.

\bibitem{wcom11}
W.~Santipach and K.~Mamat, ``Tree-structured random vector quantization for
  limited-feedback wireless channels,'' \emph{{IEEE} Trans. Wireless Commun.},
  vol.~10, no.~9, pp. 3012--3019, Sep. 2011.

\bibitem{ryan09}
D.~J. Ryan, I.~V.~L. Clarkson, I.~B. Collings, D.~Guo, and M.~L. Honig, ``{QAM}
  and {PSK} codebooks for limited feedback {MIMO} beamforming,'' \emph{{IEEE}
  Trans. Commun.}, vol.~57, no.~4, pp. 1184--1196, April 2009.

\bibitem{train10}
W.~Santipach and M.~L. Honig, ``Optimization of training and feedback overhead
  for beamforming over block fading channels,'' \emph{{IEEE} Trans. Inf.
  Theory}, vol.~56, no.~12, pp. 6103--6115, Dec. 2010.

\bibitem{ma09}
Y.~Ma, D.~Zhang, A.~Leith, and Z.~Wang, ``Error performance of transmit
  beamforming with delayed and limited feedback,'' \emph{{IEEE} Trans. Wireless
  Commun.}, vol.~8, no.~3, pp. 1164--1170, Mar. 2009.

\bibitem{mondal06}
B.~Mondal and R.~W. Heath, Jr., ``Channel adaptive quantization for limited
  feedback {MIMO} beamforming systems,'' \emph{{IEEE} Trans. Signal Process.},
  vol.~54, no.~12, pp. 4717--4729, Dec. 2006.

\bibitem{huang09}
K.~Huang, R.~W. Heath, Jr., and J.~G. Andrews, ``Limited feedback beamforming
  over temporally-correlated channels,'' \emph{{IEEE} Trans. Signal Process.},
  vol.~57, no.~5, pp. 1--18, May 2009.

\bibitem{kim11}
T.~Kim, D.~J. Love, and B.~Clerckx, ``Does frequent low resolution feedback
  outperform infrequent high resolution feedback for multiple antenna
  beamforming systems?'' \emph{{IEEE} Trans. Signal Process.}, vol.~59, no.~4,
  pp. 1654--1669, Apr. 2011.

\bibitem{osmane13}
A.~Osmane and H.~Khanfir, ``Optimal feedback updating period for coordinated
  multi-point transmission schemes,'' in \emph{Proc. IEEE Int. Symp. on
  Personal, Indoor, and Mobile Radio Commun. (PIMRC)}, London, UK, Sep. 2013,
  pp. 2281--2285.

\bibitem{zhang12}
L.~Zhang, L.~Song, M.~Ma, and B.~Jiao, ``On the minimum differential feedback
  for time-correlated {MIMO} {R}ayleigh block-fading channels,'' \emph{{IEEE}
  Trans. Commun.}, vol.~60, no.~2, pp. 411--420, Feb. 2012.

\bibitem{kim11dif}
T.~Kim, D.~J. Love, and B.~Clerckx, ``{MIMO} systems with limited rate
  differential feedback in slowly varying channels,'' \emph{{IEEE} Trans.
  Commun.}, vol.~59, no.~4, pp. 1175--1189, April 2011.

\bibitem{medra15}
A.~Medra and T.~N. Davidson, ``Incremental {G}rassmannian feedback schemes for
  multi-user {MIMO} systems,'' \emph{{IEEE} Trans. Signal Process.}, vol.~63,
  no.~5, pp. 1130--1143, March 2015.

\bibitem{chen17}
H.~C. Chen and Y.~P. Lin, ``Differential feedback of geometrical mean
  decomposition precoder for time-correlated {MIMO} systems,'' \emph{{IEEE}
  Trans. Signal Process.}, vol.~65, no.~14, pp. 3833--3845, July 2017.

\bibitem{xu14}
J.~Xu, F.~Huang, and D.~Ben, ``Quantised innovation {K}alman filter:
  {P}erformance analysis and design of quantised level,'' \emph{IET Signal
  Processing}, vol.~8, no.~7, pp. 759--773, Sep. 2014.

\bibitem{mehanna14}
O.~Mehanna and N.~D. Sidiropoulos, ``Channel tracking and transmit beamforming
  with frugal feedback,'' \emph{{IEEE} Trans. Signal Process.}, vol.~6, no.~24,
  pp. 6402--6413, Apr. 2014.

\bibitem{noh16}
S.~Noh, M.~D. Zoltowski, and D.~J. Love, ``Training sequence design for
  feedback assisted hybrid beamforming in massive {MIMO} systems,''
  \emph{{IEEE} Trans. Commun.}, vol.~64, no.~1, pp. 187--200, Jan. 2016.

\bibitem{tc15}
K.~Mamat and W.~Santipach, ``On transmit beamforming for {MISO-OFDM} channels
  with finite-rate feedback,'' \emph{{IEEE} Trans. Commun.}, vol.~63, no.~11,
  pp. 4202--4213, Nov. 2015.

\bibitem{commag04}
D.~J. Love, R.~W. Heath, Jr., W.~Santipach, and M.~L. Honig, ``What is the
  value of limited feedback for {MIMO} channels?'' \emph{{IEEE} Commun. Mag.},
  vol.~42, no.~10, pp. 54--59, Oct. 2004.

\bibitem{veeravalli13}
V.~Raghavan and V.~V. Veeravalli, ``Ensemble properties of {RVQ}-based
  limited-feedback beamforming codebooks,'' \emph{{IEEE} Trans. Inf. Theory},
  vol.~59, no.~12, pp. 8224--8249, Dec. 2013.

\bibitem{zhao_pimrc05}
Y.~Zhao, M.~Zhao, L.~Xiao, and J.~Wang, ``Capacity of time-varying {R}ayleigh
  fading {MIMO} channels,'' in \emph{Proc. IEEE Int. Symp. on Personal, Indoor
  and Mobile Radio Commun. (PIMRC)}, Berlin, Germany, Sep. 2005, pp. 547--551.

\bibitem{peel07}
C.~B. Peel and A.~L. Swindlehurst, ``Throughput-optimal training for a
  time-varying multi-antenna channel,'' \emph{{IEEE} Trans. Wireless Commun.},
  vol.~6, no.~9, pp. 3364--3373, Sep. 2007.

\bibitem{jake93}
W.~C. Jake, \emph{Microwave Mobile Communication}.\hskip 1em plus 0.5em minus
  0.4em\relax New York, NY, USA: IEEE Press, 1993.

\bibitem{dai09}
W.~Dai, Y.~Liu, and B.~Rider, ``The effect of finite rate feedback on {CDMA}
  signature optimization and {MIMO} beamforming vector selection,''
  \emph{{IEEE} Trans. Inf. Theory}, vol.~55, no.~8, pp. 3651--3669, Aug. 2009.

\bibitem{zheng99}
W.~X. Zheng, ``A least-squares based method for autoregressive signals in the
  presence of noise,'' \emph{{IEEE} Trans. Circuits Syst. {II}}, vol.~46,
  no.~1, pp. 81--85, Jan. 1999.

\bibitem{tulino04}
\BIBentryALTinterwordspacing
A.~M. Tulino and S.~Verd\'{u}, ``Random matrix theory and wireless
  communications,'' \emph{Foundations and Trends in Communications and
  Information Theory}, vol.~1, no.~1, pp. 1--182, 2004. [Online]. Available:
  \url{http://dx.doi.org/10.1561/0100000001}
\BIBentrySTDinterwordspacing

\bibitem{lloyd82}
S.~P. Lloyd, ``Least squares quantization in {PCM},'' \emph{{IEEE} Trans. Inf.
  Theory}, vol.~28, no.~2, pp. 129--136, Mar. 1982.

\bibitem{fisher39}
R.~A. Fisher, ``The sampling distribution of some statistics obtained from
  non-linear equations,'' \emph{Annals of Eugenics}, vol.~9, pp. 238--249,
  1939.

\bibitem{mukkavilli03}
K.~K. Mukkavilli, A.~Sabharwal, E.~Erkip, and B.~Aazhang, ``On beamforming with
  finite rate feedback in multiple antenna systems,'' \emph{{IEEE} Trans. Inf.
  Theory}, vol.~49, no.~10, pp. 2562--2579, Oct. 2003.

\end{thebibliography}

\end{document}